# TRUSTWORTHY 100-YEAR DIGITAL OBJECTS

## DURABLE ENCODING FOR WHEN IT'S TOO LATE TO ASK

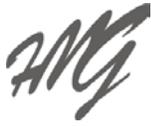
**H.M. Gladney**
HMG Consulting
Saratoga, CA 95070

**R.A. Lorie**
IBM Almaden Research Center
San Jose, CA 95120

How can an author store digital information so that it will be reliably useful, even years later when he is no longer available to answer questions? Methods that *might* work are not good enough; what is preserved today should be reliably useful whenever someone wants it. Prior proposals fail because they confound saved data with irrelevant details of today's information technology—details that are difficult to define, extract, and save completely and accurately.

We use a virtual machine to represent and eventually to render any data whatsoever. We focus on a case of intermediate difficulty—an executable procedure—and identify a variant for every other data type.

This solution might be more elaborate than needed to render some text, image, audio, or video data. Simple data can be preserved as representations using well-known standards. We sketch practical methods for files ranging from simple structures to those containing computer programs, treating simple cases here and deferring complex cases for future work. Enough of the complete solution is known to enable practical aggressive preservation programs today.

## 1 INTRODUCTION

Preserving digital data has received increasing attention since 1996. Some authors deem the technical challenges difficult. [Granger] There has been little substantial progress:

> "Preservation of digital information is complex because of the dependency digital information has on its technical environment. … digital resources present more difficult problems than conventional analog media such as paper-based books. … there is a lack of proven preservation methods to ensure that the information will continue to be readable."  [Lee]

From the perspective of a computer scientist in 1995, long-term digital preservation faced two conceptual challenges for each saved document. We here address the first: how to save a digital entity so that its content will be as useful a century from now as it is today. This challenge is the kind of language problem that has been central to computer science since it emerged as a discipline in the 1960's. Its core can be restated as, "ensure that an arbitrary computer program will execute correctly on a machine whose architecture is unknown when the program is saved."

We intend to enable unambiguous communication with people with whom dialog is impossible, and to avoid any restriction of the content that today's authors can communicate. For this, we need language that we can confidently expect our descendants to understand. In effect, we are proposing a set of durably useful information representations—**durable encoding**.

The core idea is to express the difficult parts of any information to be preserved in terms of programs of simple virtual computers, following a [Lorie] proposal and elaborating that to deal with what seem the most difficult cases—programs with multitasking and real-time dependencies, and that further include real-world input and output, such as that required for robots. [Lorie] is couched in terms of a "universal virtual computer" (UVC) that is computationally equivalent to a Turing machine. The current paper elaborates [Lorie] towards practical implementation for cases of intermediate difficulty, and suggests how to extend this to the most difficult cases, but leaves this further elaboration to future work.

## 1.1 Overview of *Trustworthy 100-Year Digital Objects*

As illustrated by the [U.S.] National Digital Information Infrastructure Preservation Program plan [LC NDIIPP] and many published articles, comprehensive digital preservation demands attention to many aspects—conceptual, organizational, and technical. The scope of the task should not be surprising, because what is at issue is future access to most intellectual work.

The *Trustworthy 100-Year Digital Objects* work treats only the technical component of digital preservation. This includes only those digital library technology aspects needed to counter technology obsolescence and fading human memory. It necessarily exploits the teachings of epistemology in addition to those of basic computer science.

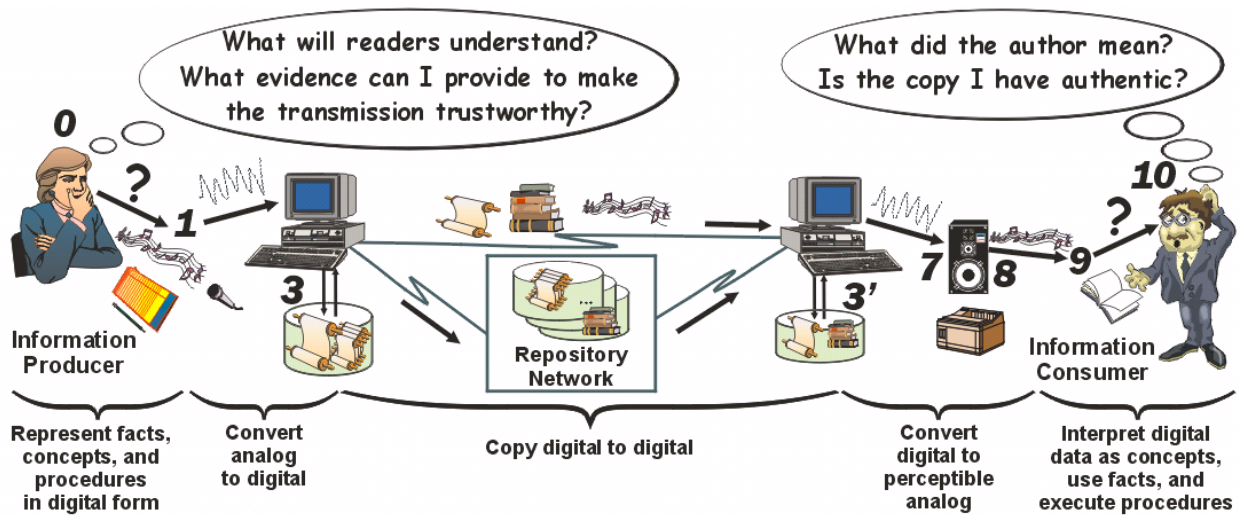

**Figure 1: Document flow from an author to a reader.**
The digital object numbering corresponds to that in other *Trustworthy* … papers.

Consider what someone a century from now might want of information stored today. Figure 1 helps us discuss communication reliability challenges. Since eventual users of preserved information might suffer harm or loss if they are misled, we pay attention to the potential distortions in the channel that transmits an input **1 to become a replica 9**. The figure suggests the technical challenges of digital preservation—finding, demonstrating, and testing methods for:

- Ensuring that a copy of every preserved document survives as long as it might interest someone;
- Ensuring that consumers can find and use any preserved document as its producers intended;
- Ensuring that any consumer has access to information for deciding whether a preserved document is sufficiently trustworthy for his intended application;
- Hiding information technology complexity from end users (producers, archivists, and consumers); and
- Replacing human effort for preservation by automatic procedures whenever doing so is feasible.

This list, in turn, suggests technical topics that can be handled almost independently of one another.

I. Repositories that store packaged works, and that provide search and access services whereby information consumers can find and obtain what interests them. Preservation requires no digital library technology beyond what many papers describe and software offerings already provide.

II. Replication mechanisms that protect against the loss of the last remaining copy of any work. [Reich] teaches a satisfactory solution.

III. Some number of socially-communicated languages and standards that are not themselves parts of the technical solution, but that are essential starting points.

IV. A method for packaging a work together with metadata that includes provenance documentation and reliable linking of related works, ontologies, and rendering software[1] and of the pieces of each package with one another. This topic has already been addressed in [#3].
V. Topic-specific ontologies provided and maintained by academic and other professional communities.
VI. A bit-string-encoding scheme to represent each content piece in language insensitive to irrelevant and ephemeral aspects of its current environment. This is the topic of the current paper.

*Evidence After Every Witness is Dead* [Gladney 3] addresses document authenticity evidence by describing the structure and use of each archived object. This is realized in the metadata and structuring XML portions suggested by Figure 3. The archived object design includes the following key elements.

(i) Each TDO contains its own world-wide eternal and unique identifier and its own provenance metadata, and is cryptographically sealed;
(ii) References to external objects are also sealed together with the identifiers of their referents;
(iii) A network of certification keys is grounded in published and frequently changed keys of trustworthy institutions. Final sealing of a preserved document by such an institution creates durable evidence of its deposit date.

The current paper teaches a method of encoding any kind of data whatsoever to be forever useful. This method would be applied to most kinds of payload bit-string called for in Figure 3. Its key ideas include:

(iv) That we can and must enable information producers to separate irrelevant environmental information from information essential to each producer's intentions, encoding only this essential information.
(v) That extended Turing-complete virtual machines can represent anything that can be written;
(vi) And that such machines can themselves be described completely and unambiguously.

Figure 1**Error! Reference source not found.** reminds us that an author typically tries to encode information so that an eventual reader has a good chance of comprehending what the author intends. In one kind of ideal scenario, perfect performance would be characterized by the consumer understanding exactly what the producer intended to communicate. In addition to human imperfections of authors and readers, the first and last steps—the steps labeled with question marks—have unavoidable imperfection originating in limitations of language. These limitations can be characterized as failures of communication that occur because jargon, expectations, world views, and ontologies are at best imperfectly shared, in the sense of *I cannot tell you what I mean* and *I cannot know how you interpret what I say*. Discussion of such difficulties is delayed to [#4] and [#5], which discuss the limits of what computers can do and what these theoretical limitations teach about managing issues of information authenticity.

*Syntax and Semantics—Tension between Facts and Values* [Gladney 4] provides epistemological arguments justifying that the methods of the current paper and [#2] do as much as mechanical methods theoretically can do towards preserving digital information, and that they attempt no more.

*What's Meant? Intentional and Accidental in Documents* [Gladney 5] examines what information producers can do to minimize eventual readers' misinterpretations, given that communication invariably confounds what it intends to convey with accidental information.

Thus, the current article and [#3] treat only the communications and transformations suggested by the arrows **Error! Reference source not found.**, and not even all of those. Specifically, they treat only what can be accomplished with digital technology, deferring conversion from and to analog representations for future attention. Treating input/output (I/O) directly is also deferred, as a huge digital corpus can be preserved without encountering the complexities of I/O or other tricky topics. Happily, based on what we describe now, practical progress is possible, allowing the preservation community to delay working on more complex file formats and other tricky topics only when they become urgent.

---

[1] We use 'render', a synonym for 'interpret', for the special case in which the outputs are intended to be useful to human beings without further translation.

## 1.2 Overview of the Current Article

We explain as simply as we can how to store today's information to be as useful for our descendants as it is for our contemporaries. We treat a single digital entity, not here addressing questions of digital collections or repository design. We further limit the treatment to static data objects and to the simplest computer programs—a class called filters, but speculate how we might treat more complex programs.[2]

§2.2 reviews earlier attempts to preserve information, explaining why critics have concluded that these methods are unsatisfactory. §2.3, showing that using a virtual computer definition can succeed, sets the stage for describing durable encoding, how to apply it, and why it is sure to work.

For brevity, we call any suitable virtual computer a "Universal Virtual Computer (UVC)". We also use the year 2005 as a surrogate for "in the near future", and the year 2105 as a surrogate for "too long from now for any conversation between content consumers and content producers." Similarly, "M2005" denotes a current computer and "M2105" denotes its eventual successor. We use "DO", whose mnemonic is "digital object", to denote information packaged as a single blob (binary large object) for interchange, but not necessarily prepared for preservation. Typically, a DO would contain several blobs—such as the encodings of multimedia information—and associated metadata, all packaged with administrative information and tabulated relationships among DO parts and with other DOs. Finally, we use "TDO", whose mnemonic is "trustworthy digital object", to denote a DO encoded and packaged for long-term preservation; this jargon will help distinguish durable encoding from seemingly similar proposals.

The arrows from 1 until 9 in Figure 1**Error! Reference source not found.** depict purely syntactic transformations that permit only three tactics. We can avoid introducing irrelevant information (such as details about today's computing environment). Secondly, we can transform information from representations that seem difficult to representations that we believe consumers will be able to understand because of shared cultural contexts and stable international standards. Finally, we can sometimes provide redundant representations that reduce misunderstanding between information producers and consumers.

§3 sketches our technical proposal for transforming today's information representations into representations likely to be intelligible a century from now, doing so in enough detail to explain how language translations support each TDO instance. §3.2-§3.4 explains as simply as possible what [Lorie] specified in more technical terms—how to exploit a Turing-complete virtual machine for durable encoding.

The technical solution presented here and in [#3] exposes conceptual problems that influence responsibility assignments among information content producers, custodians, and consumers. We identify one such issue—the boundary between essential and accidental information, but defer discussing it to [#5].

## 1.3 What is Needed to Keep Information Useful?

We assume that a DO survives into the distant future by happenstance or by the care of some repository, and is available to some consumers. What can we do now to ensure that any of these consumers might be able to exploit the DO as its producer(s) intended when they saved it? In other words, what conditions on a DO make it also a TDO?

Information must be **fixed** if it is to be preserved. Here "fixed" is intended precisely as it is intended in copyright law. [Nimmer] Any information we want to save can therefore be represented as a set of computer files—a set of bit-strings. This includes dynamic behavior, which would be represented by programs, and perhaps also by snapshots or sequences of states of execution.

We intuitively want to convey something along the lines of, "At such and such a time, B (a person) communicated a record X (a manuscript, painting, musical performance, table of facts, computer program, or any other information of interest to others)." To make such a record acceptable to archivists, it must contain provenance information, with the whole being protected against undetected tampering. [#3]

In digital libraries, a digital entity is registered under a certain name, a bit-string copy of the entity is stored in a safe place, and an entity description is saved in a database that also points at the stored bit-string. This pattern mimics the organization of ordinary libraries. However, unlike what happens with a physical

---

[2] A filter is a program whose input is some number of files or bit-strings, whose output is some number of files or bit-strings, and that has no side-effects (makes no changes to the state of the computing machine or the world.)

Page 4 of 20© 2004, H.M. Gladney    C:\staging\41765454-3808-203F85\in\41765454-3808-203F85.doc

book, the consumer cannot extract the information without a computer and, seeing it, would not know its meaning, since his schooling probably did not include the bit-string language.

Suppose that in 2005 we create digital information, saving this on a removable disk called D2005. Suppose further that, a century later, somebody (an information consumer) wants to use this information. What must happen to make this possible and effective? Four things must be true in 2105

**Condition 1**: A volume D2105 containing bit-strings that had been on D2005 must be found.
**Condition 2**: D2105 must be sufficiently intact to be read.
**Condition 3**: A computer M2105 must be available to read the raw bit-strings from D2105.
**Condition 4**: M2105 must render the bit-string, making it comprehensible to the consumer.

**Condition 1**: Engineers predict long lifetimes for data inscribed on certain media such as single-crystal nickel, but these media are today impractical for large data collections. Suppose D2005 endures only for N years; what is needed to support consumers in (N+100) years or later? However large N is made, the information will be lost unless somebody acts soon enough, copying at-risk bit-strings to new media.

**Condition 2**: Here, "sufficiently" hints at the complexities of error correction codes and other recovery techniques from damaged media.

**Condition 3**: Obsolete computers are very difficult to keep in working order. Typically, they "die" in about M years, with M less than N. This can be solved by moving bit-strings onto media accessible by some then-modern computer, doing this before computers of the previous generation become inoperable.

Satisfying these three conditions will ensure that any bit-string saved today will be readable by some computer whenever it is wanted. However, according to **condition 4**, one must still render each bit-string to expose its useful meaning. It is this challenge that the current article addresses.

## 2 TECHNICAL ALTERNATIVES

What methods might be used for making a bit-string durably useful? Investigators are finding that straightforward extensions of institutional practices developed over the past century are insufficient. [Marcum] Apart from durable encoding as proposed in this article, we know of only the following possibilities. Why have these prior proposals not succeeded?

- **Natural language** contains too many ambiguities to be used alone for precise communication. Furthermore, natural language changes too rapidly for unqualified confidence in it as a preservation foundation.
- **Formal semantics** [Bjørner], such as denotational semantics [Gordon], might mitigate weaknesses of natural language, except they seem to be understood only by language theorists. However, the Backus-Naur Form (BNF) is sufficiently simple and widely taught for preserving **syntactic** rules.
- **Standards** expressed in a combination of natural and formal (mathematical) languages are essential, but are practical only for simple data types. (These include essential starting points; see §2.1.)
- **Transformative migration** has been carefully considered for almost a decade. We'll explain why we agree with an emerging consensus that this approach is inadequate.
- **Preservation emulation** has been considered as an alternative to migration, but seems inadequate for the same reason. Both try to preserve obsolete technical environments—information that is both difficult to capture correctly and also irrelevant to authors' objectives. [#5]
- **Multiple formats** [Gilheany] combined with "**digital archeology**" might be practical for relatively simple data formats. "Digital archeology" [Ross 99] is an almost "do nothing" tactic that leaves most of the work to our descendants. Future readers' problems can be eased by the Rosetta Stone lesson; it was the redundancy of multiple languages expressing the same information that enabled Young and Champollion to decipher the Egyptian hieroglyphic language. [Donoughue]

"Digital archeology" reminds us that that extra work today for preservation is not always essential—merely cost-effective. It leaves rescuing content to (agents for) whoever wants what it conveys. This might be done as a matter of policy, or simply because we cannot persuade people to prepare works for preservation. Choosing to prepare a document for retention is an economic decision that depends on the ex-pected number of retrievals, the cost of preparation, and whether one is willing to expend on behalf of unknown future beneficiaries. In contrast, preserving computer programs cannot rely on digital archeology because programs contain insufficient redundancy; a single bit error can cause serious failures.

## 2.1 Standards for Encoding

**Digital preservation needs only small extensions over information interchange and digital library technologies that are already widely deployed.** Most of the software technology needed is already available, or soon will be, and is being standardized. This includes multimedia encodings (e.g., [MPEG]), XML syntax [XML], semantics encoding [RDF], cryptographic tools, and encodings of scientific data.

> "[Yale Electronic Archive] team members maintain that standards, despite their flaws, represent an essential component of any coherent preservation strategy adopted.
>
> "… standards mean durability. Adhering to commonly and widely recognized data standards will create records in a form that lends itself to adaptation as technologies change. … Identifying those standards has an element of risk about it, … but at the moment some of the choices seem fairly clear.
>
> "… The right standards are the ones that allow transformation into as many forms as the present and foreseeable user could wish. … The XML suite … is most desirable because it is portable, extensible, and … can generate everything from ASCII to HTML to PDF and beyond." [Okerson, pp.15-17]

We might want structures and mechanisms that admit no misunderstandings. However, there is no proper starting point for defining any language—whatever language one might choose as the starting point cannot itself be defined without language. The best we can do is choose formal and natural language starting points for simple data types, admit that we cannot fully avoid risks of misunderstanding, and use heuristics and redundancy to provide hints that might deflect misunderstandings.

To establish starting points for encoding hierarchies of increasingly complex data types, our main concern needs to be for primary syntactic and semantic encoding standards, rather than secondary standards that can be defined with these primary standards. We can choose primary standards conservatively. The eventual consensus choices will be community decisions that depend on deliberations with far more detail than the current article can include. Happily, the standards needed for durable encoding are limited to ASCII, Unicode UTF-8, and core portions of XML.

We regard as secondary XML schema such as the Metadata Encoding and Transmission Standard being developed by the Digital Library Federation [METS]. We can regard RDF and other semantics standards as secondary because we can specify them with XML and mathematics.

## 2.2 "Transformative Migration" and "Preservation Emulation"

It seems to be common opinion that there are just two possibilities for making complex content durably intelligible: transformative migration and preservation emulation.

The idea of "transformative migration" (a.k.a. "active migration") is to transform saved data to encoding with replacement technology whenever the imminent demise of some technology threatens their interpretability. If programming technology continues to change as quickly as has been the case in the most recent two decades, this could be needed roughly every five years. [Mellor]

The idea of "preservation emulation" is to accompany stored content by description of all technical factors that might affect interpretation—a "complete" characterization of the hardware and software environments within which the content is used. When the content becomes of interest, an emulator for M2005 executing on M2105 would be constructed to enable execution of 2005 programs. [Rothenberg]

For about seven years, these two approaches have dominated discussion of representing complex data for preservation. The debates have been extensive, but have not resolved the issues. To help us choose a course of action, it is sufficient to note that **neither method has been demonstrated to permit a sure way of avoiding syntactic errors.** What they fail to provide is sure conversion to future interpretable languages for computer programs or for other information for which broadly accepted data standards are insufficiently reliable.

When was the last time you saw and understood the specification of a practical computer or program more than a few hundred lines in length that was complete and known to be correct? We have never seen one, and have never **fully** understood whatever documentation was supplied. We need not repeat the arguments about migration and emulation, but can quote passages that typify the debate.

> "Rothenberg does make an insightful and enterprising case for … emulation as the "true" answer to the digital longevity problem. … [He] logically argues that once a hardware system is emulated, all else just naturally follows. The operating system designed to run on the hardware works and software application(s) that were written for the operating system also just work. Consequently the digital object behaves and interacts with the software as originally designed.
>
> "However, emulation cannot escape standards. … If the manufacturer of a piece of hardware did not adhere 100% to the standard, then the emulation will reflect that imperfection or flaw. Consequently, there is never a true solution … [so] that a generalized specification for an emulator … can be constructed."
>
> [Okerson, pp.15-17]

> "… Rothenberg is fundamentally trying to preserve the wrong thing by preserving information systems functionality rather than records. As a consequence, the emulation solution would not preserve electronic records as evidence even if it could be made to work and is serious overkill for most electronic documents where preserving evidence is not a requirement. …
>
> "[He] asserts that "migration is labor-intensive, time-consuming, expensive, error-prone, and fraught with danger of losing or corrupting information" and that "automatic conversion is rarely possible". …
>
> "Rothenberg's abstract argument, that translation always involves loss of information, is plausible, but not … very relevant. If it was true, his own case for emulation … would be fatally flawed. But for the case to be validated, Rothenberg would need to specify just what characteristics of records are crucial to preserve (or … their properties as evidence) and how these are affected by "translation"." [Bearman]

The weaknesses described have not been addressed. The arguments have not been refuted.

## 2.3  How Durable Encoding is Different

Durable encoding avoids the problems sketched in §2.2 by representing difficult content types with the aid of programs written in UVC code (§3.5). Since this involves language translations similar to those encountered throughout computer science, we can implement it using well-known techniques. Since the UVC design can be very simple (compared to designs practical for hardware implementation), it can be specified completely and unambiguously for future interpretation.

A 1970's *compiler* (e.g., a COBOL compiler) was a computer program that accepted a file in some source language (e.g., COBOL) and produced a new file in some target language, which might be the *machine language* for the hardware on which the program was to execute. Its output conformed to *principles of operation* that describe the *instruction set* of some computer. The core of any compiler is a parser that decomposes input into source language primitives—phrases from a constrained vocabulary. The rest is a target language generation subroutine for each kind of source language phrase; each such subroutine produces part of an output bit-string. The compiler itself is a program in the code of some computer, which might or might not be the target for execution of the compiler's output programs.

A programming language *interpreter* differs mainly in that, instead of producing an output string of target machine instructions, it immediately executes each instruction a compiler might have written to an output file. An interpreter must execute on the computer for which it produces instructions.

Modern compilers are more complicated to facilitate translation for otherwise incompatible hardware and to permit late binding of resources, such as data files and drivers for otherwise incompatible printers and display devices. To reduce the cost of supporting incompatible instruction sets, the compilation is partitioned, with the compiler "front end" producing code for a virtual computer (such as Sun Microsystem's Java Virtual Machine (JVM)™) and with, for each target machine, a "back end" compiler or interpreter that translates resp. executes the virtual machine instructions. To support late binding of input sources and output targets, the compiler produces subroutine call stubs with free variables and a symbol table of these variables and constraints on the values that can be bound to each variable. Each stub calls a "driver" program interface that is generic for a service class (e.g., printers). At the time the compiled program is invoked, or during its execution, an opportunity is provided to choose or change the specific resource bound to each variable encountered.

To implement durable encoding, we change these well-known practices slightly. In 2005, we write programs in UVC code that will be translated in 2105. Translating will be easier than writing most compilers.

Many compiler complexities are irrelevant, because each input will be static (not being changed by programmers) and because it can be assumed to be without error.

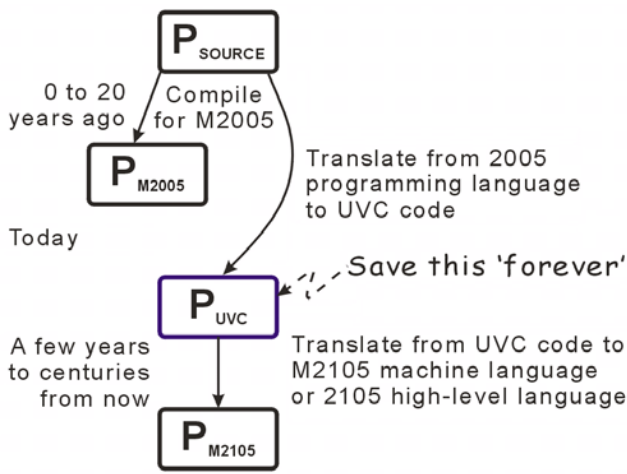

**Figure 2: Translations of a program—from source into future machine code**

Figure 2 suggests the history of a program **P** that is to be preserved. For each data type, some programmer must create a UVC program, either by writing a rendering program from scratch or by translating a current program to UVC language. Each subscript denotes a language in which **P** is encoded (represented). Each of $P_{SOURCE}$, $P_{M2005}$, $P_{UVC}$, and $P_{M2105}$ expresses precisely the same program in a different language. $P_{SOURCE}$ will have been written a few years ago in a language such as COBOL, Fortran, or C, and translated (compiled) into M2005 machine language. $P_{M2005}$ executes on a current computer, and $P_{M2105}$ will execute on a computer whose specification is not yet available.

We do not translate $P_{M2005}$ directly to $P_{M2105}$ because reliable M2005 and M2105 instances might never be available simultaneously. Even if some 2105 programmer believed he possessed a good M2005 specification, he might be unable to test that his $P_{M2105}$ executed precisely the way $P_{M2005}$ had executed, because no working M2005 machine would be available.

Instead we create and save a new encoding, $P_{UVC}$. This works because our UVC (§3.5) is simple enough for us to write its complete specification, and to test that any UVC implementation is entirely correct. We can also test that a UVC program correctly emulates some other program that it is intended to replace, because both programs will be concurrently available. This prescription bypasses technical dependencies on programming languages, because whoever manipulates the supported data types will have a $P_{M2005}$ copy. Intellectual property issues, such as those associated with proprietary intermediate language specifications, remain a challenge.

# 3 DESIGN OF A DURABLE ENCODING ENVIRONMENT

Every kind of persistent digital object will have a linear (one-dimensional) encoding. Operating systems or application programs provide for converting this linear form to other forms and for linearizing from other forms. We therefore treat only linear encodings (blobs a.k.a. files). Furthermore, without reducing generality, we can treat the files as bit-strings in which any bit can be accessed independently of other bits.

One might save many data types with existing or expected electronic data processing (EDP) standards. When this proves unsatisfactory for some data type because sufficiently simple and reliable EDP standards are not available, we supplement the content by an accompanying UVC program.

We focus on preserving computer programs, because they include the most difficult preservation challenges and because we intend to handle all kinds of information. For instance, engineering and military simulators, as well as computer games, are part of the cultural record. Furthermore, to communicate the intent of complex data we can save 2005 programs that render them in as many ways as we guess might be helpful to some 2105 consumer who wants to understand or otherwise use what's saved. Programs are a good target for attention because:

- ➢ A programming language specification is often more complex than programs in that language.
- ➢ The programs we will want to save will have been tested.  In contrast, a programming language specification might be incomplete or incorrect.  Only rare and expensive experts can determine its quality.  Even for them, doing so is likely to be time-consuming and tedious.
- ➢ Many programs have little internal redundancy with which errors might be detected.
- ➢ An error in program translation can precipitate drastically incorrect behavior in its executions.

## 3.1 Packaging Content and Metadata

> "Metadata is a central element of any model designed to ensure that preserved data is functional. … the complexity of inter-relationships between resources and the various software applications used to run them, may be easily overlooked when creating the metadata elements of the wrapper."  [Ross 00]

TDOs are packaged with XML and Unicode, as is independently proposed by many authors. [Tennant] We accept the growing consensus for packaging closely related data blobs into a single container that carries metadata, links to prerequisite documents, and structuring information in XML.  This is suggested by Figure 3, which [#3, §3] elaborates.  We can save any digital object set **Y** within a **TDO** instance.  Any blobs with important relationships to **Y**, and that we do not want in the same **TDO** package, can be safely linked to other **TDO**s that we create to hold them.

Page 9 of 20© 2004, H.M. Gladney    C:\staging\41765454-3808-203F85\in\41765454-3808-203F85.doc

[Figure 3 diagram: shows XML metadata code at top (labeled "Metadata"), followed by three boxes labeled "Payload blob #1 (e.g., a computer program)", "Payload blob #3 (e.g., a first data blob)", and "Payload blob #3 (e.g., a second data blob)", with XML code between boxes (labeled "Structuring XML")]

**Figure 3: Structure of an exemplary TDO (a trustworthy DO).** Whether related blobs are packaged together or in separate TDOs is decided by some information producer. [#3] teaches how to construct secure links between TDOs whose correct relationships are essential to the information producers' objectives.

The schema and standards needed by most TDOs can be handled by references to distinct TDOs. XML and Unicode UTF-8 are sufficient for TDO packaging and metadata. Fraudulent substitution of linked references would create security exposures for sensitive information that are reliably detectable by a scheme described in [#3]. What remains is to specify how to communicate each embedded payload blob (Figure 3). We distinguish methods for cases of increasing difficulty.

**Case 1**: The blob is readily understandable by a human reader without specialized expertise. The only practical examples are text encoded in a well-known alphabet, such as ASCII, and binary raster images represented without compression. These can be saved without encoding beyond text declarations of their types and dimensions.

**Case 2**: The data are too complex for human beings to understand without machine assistance. For this, some 2105 programmer must write a decoding program, using a 2005 program specification in natural language conveyed in some TDO referred to. This program must be able to depict for a (human) consumer the interpretation of every important DO aspect. A line drawing might be conveyed this way, because its presentation algorithm is simple and amenable to a natural language description.

**Case 3**: The content is too complex for reliable **Case 2** treatment. This includes scientific tables, vector graphics data, and also representations of text created by word processing programs. Convey-ing one or more rendering programs is the only way to ensure that the consumer will be able to comprehend the payload. Whatever language this program is written in must be sufficiently understood in 2105 for writing an interpreter or compiler.

**Case 4**: The content is a program or system to be saved for its own sake. We must ensure that this program will execute in the M2105 machine. For instance, if we want to preserve the look and feel of an Apple MAC or the user interface of a Computer Aided Design (CAD) system, we must save M2005 programs together with instructions for executing these programs in M2105.

**Case 5**: Programs with concurrency and critical timing relationships can be represented by modest extensions of the **Case 4** mechanism. Real-time input can be represented by memory locations and triggers that accept changes from outside. The most complex programs will be simulations, such as virtual reality programs and aircraft pilot trainer control programs. For instance, a pilot-training program must react to numerous cockpit controls, provide a video simulation of what the pilot would see in many dials and simulated flight path displays such as those used for landing when runways are hidden by fog, and mimic the reaction times of aircraft, even though future computers will run at different speeds than today's.

**Case 1** and **Case 2** instances present no problems needing discussion in the current article. They could be handled with **Case 3** procedures, but people might prefer simpler methods. In **Case 3** and in **Case 4**, we must save a program. The difference is that for **Case 4**, we must interpret M2005 machine language, because we cannot save program behavior reliably except by saving the program itself. We defer treating **Case 5.**

Many kinds of data to be preserved will involve more than one source file that needs to be represented as a bit-stream object in a package that binds all necessary objects together, together with labeled links between parts of the complete package. [#3] already described a way of handling this in the XML that surrounds the payload blobs (Figure 2). Briefly, a **TDO** manifest indicates the treatment needed for each object, and the metadata includes a table of relationships of blob pairs belonging together. For instance, a scientific text might be associated with graphics, and also with spreadsheets. It might be important to indicate the connection between a text paragraph and a set of spreadsheet cells.

## 3.2  Preserving Complex Data Blobs as Payload Elements

For **Case 3** we save the data itself in whatever form it was produced and/or is usually used, together with a **UVC** program that provides intelligible or otherwise useful renderings of those data. (Figure 4) Optionally, we might transform the input **Data** to some **Data'** in order to simplify the programming work. If we do this, we would also save the original **Data** in order not to block new purposes whose nature we cannot today predict**.**

The program, written in UVC code, consists of a parser and subroutines for some number of result bit-strings. More than one result string might be wanted in order to reduce unavoidable ambiguity. [#5] More than one result string might be needed to represent information not included in the single string that might otherwise be chosen. We might even decide to produce several result strings simply for consumers' convenience; for instance, we might produce a form for printing tables, a file of commands to load a relational database, and instructions for drawing a directed graph.

We can test the correctness of this UVC program by comparing the results of emulation on M2005 with those from a UVC emulator running on a computer incompatible with M2005 (such as an Apple MAC if we started with an IBM PC). A good further test would be to translate exemplary results and observe whether independent human readers understand the information that should be conveyed.

Our 2005 **UVC** program will be interpreted in 2105 by a **UVC** interpreter written to operate in a **M2105** environment and to process the saved data**.** Each of the **Data** file, the **Data'** file, and the **UVC program** is a 2005 blob that we might package into a single **TDO,** together with metadata that might be needed by the restore application execution in 2105 (Figure 3), and store this TDO on a volume **D2005** (Figure 4). This **TDO** is copied from volume to volume whenever needed to avoid losing its inherent bit-string pattern, possibly using ordinary computing service backup and replication processes and possibly with something more powerful [Reich], so that a faithful copy eventually is available on some 2105 volume **D2105**.

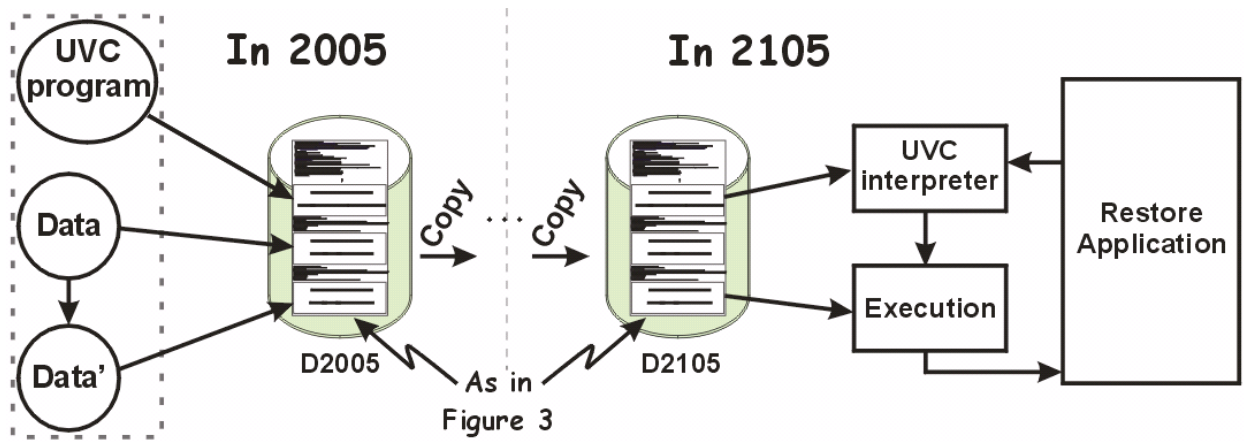

**Figure 4: Durable encoding process for complex data**

To prepare for using the preserved bit-strings our 2105 successors must write a UVC emulator that executes on some **M2105**, and create a restore application. As Figure 4 suggests and §3.5 describes, the restore program must pass the locations of the saved UVC program, saved **Data'** strings (there might be more than one), and addresses where results should be stored. These would be the call parameters for invoking the emulator. It also needs to print or otherwise handle the results.

## 3.3 Preserving Programs as Payload Elements

In **Case 4**, we save the input **Data,** the **Application program**, an **M2005** program, and an **M2005 Emulator** written as a **UVC** program. We package these 2005 objects and their relationship specifications into a **TDO**, and copy this bit-string from volume to volume over the years. (Figure 5)

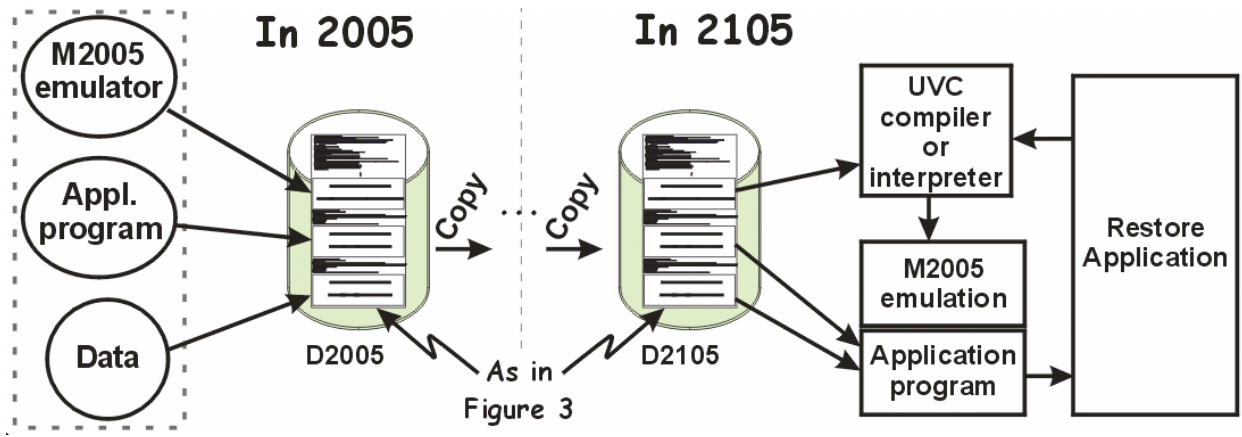

**Figure 5: Durable encoding process for preserving a program**

In 2105, a restore application uses a **UVC** interpreter written in **M2105** code to execute or compile the **M2005 Emulator.** This output executes the 2005 **Application program,** taking as input the 2005 **Data.**

## 3.4 Preserving Dynamic Objects and Databases

Modern multimedia recordings are mostly digital files that can be preserved as already sketched in §3.2. In fact, people might come to consider such recordings fit for preservation "as is" rather than using the UVC mechanism if optical disk recording standards solidify. Older recordings in analog formats include perceptible imperfections. Good technology to convert these to digital derivatives is affordable even for hobbyists, and extensive literature (e.g., [Calas], [Pohlmann]) teaches how to deal with imperfections.

Page 12 of 20© 2004, H.M. Gladney        C:\staging\41765454-3808-203F85\in\41765454-3808-203F85.doc

> The static approach to document storage is most easily followed with documents that have a permanent or long term retention period. Dynamic documents … do not fit well into this static model. [Gilheany]

This and other published comments (e.g., [Steemson], [Depocas]) suggest that parts of the interested community are perplexed about digital preservation of dynamic information, such as performances and database contents. **Dynamic behavior** is articulated **either by a prescription of what should happen**—a rule set for an event sequence, the text of a theatrical play, a musical score, a computer program, or whatever prescription is customary for the genre at hand—**or else by a history of what has happened**—a performance record, such as a business journal, a multimedia recording, or a digital log.

**A special case of a recording is a snapshot—a recording part related to a single instant, such as a photograph, a single business transaction, or part of a computer's instantaneous state. The details of what will be useful for any kind of recording are often much the same as might have been customary before digital technology was widely used.**

Database management systems (DBMSs) typically provide for snapshot and log creation, addressing administrators' needs for images of internally consistent states and database reconstruction for failure recovery. This is done using an event log that includes incomplete transactions and the ability to create a snapshot—an image of the database state at some instant, adjusted to include only transactions that have been declared complete by application processes. Snapshots can be for any instants that some agent chooses. This transaction log and set of snapshots can be processed to create an image of the database state at any past instant within the time span represented by the log and snapshots.

A variant of this DBMS service might be applied at the server end of dynamic Web pages. ([Brown] wrote, "If the utility of both the fixed and the fluid is recognized, the Web may develop much of its innovative power from the possibility of producing documents that combine both fixity and fluidity. Already, many documents retain a constant text while their links are continually changed. … This interplay between fixity and fluidity, formerly possible only on the scale of collections, may now become a central feature of individual documents.") Preserving database states, sequences in time, or performances as alluded to by [Steemson] encounters no technical challenges beyond those addressed elsewhere in the current article.

## 3.5  Universal Virtual Computer and Its Use

A single UVC definition would be enough, but nothing precludes developing a new version if this is really needed; each TDO identifies the UVC it uses. Since an **M2105** emulator for a particular UVC would be suitable for many object instances, it will be shared just as compilers are shared.

The UVC could be as simple as a Turing machine. This choice would make UVC programming tedious. More practical is a machine with generous memory and register structures. Our prototype suggests that it will cost less than a man-year to create an emulator for such a virtual computer. Of course, a UVC emulator for M2105 would work for any saved information; i.e., it needs to be written only once for each future machine architecture. The incomplete UVC description (for more detail see [Lorie]) that follows is just enough to suggest how 2005 and 2105 programmers would accomplish their responsibilities.

- The memory has an unbounded number of segments, each of unbounded length. Thus, its programmers need be little concerned with data sizes.
- The memory is bit-addressable. A typical reference is to a bit offset from the origin and a bit length. Thus, the UVC is not biased towards any particular byte length.
- The machine has an unlimited number of registers, each of unlimited length, supporting any number of variables of any lengths.
- Fewer than 40 UVC instruction definitions provide the usual kinds of copying between registers and memory segments, integer arithmetic, and tests comparing values in two registers.
- For flow control the UVC includes a branch operation. A UVC programmer needs to create loops and other high level flows "by hand", as we did before high-level programming languages were invented.
- This prototype UVC supports only a single programming thread. More complex examples than we are prototyping would need UVC instructions that read a clock, handle interrupts, fork to implement multiprogramming, and provide latches for synchronization.

> In its early version, the UVC did not have any I/O operations because it was easier to simply have the Restore program read the data and transfer it to the emulator, which knew where to store it in the UVC memory. A very simple I/O mechanism has now been added, in anticipation of more complex applications. (Figure 4 and Figure 5)

UVC documentation, application program documentation, and input and output descriptors, are ASCII and BNF files that include:

**Item A:** A natural language description of the alphabet(s) used in **B**, **C**, **D**, and **E**.
**Item B:** A natural language description of each program, and its inputs and outputs.
**Item C:** A UVC program **U** that might be a parser. In this case, for each input token, a terminal parse tree node would contain a result method.
**Item D:** Schema for the input and output bit-strings of **U**.
**Item E:** A description of the invocation and return sequences of **U**. This might specify into which address segment the restore program needs to load the data, in which memory segments it will find the various renderings provided by **U**, and how to print these results or pass them to other 2105 programs that would use them as their 2005 producers intended.

A restore application (see Figure 4, which is drawn for the case of a single data bit-string) would read the UVC program U and each input bit-string into a different memory segment (without any transformations) and then call the emulator. That would execute U, which would put its results into some number of different segments, and call back to the restore application either many times (for a UVC interpreter) or when complete (for a UVC compiler). The restore application can be written to handle results as the 2105 programmer wants, using access methods described by the 2005 UVC programmer.

# 4 DISCUSSION

> "Preservation and management of many types of digital information require transformation of the original data to new formats or canonical forms. Research is needed to better define and characterize transformation processes so that they can be automated, and so that transformations made on the original data can be documented."     [NDIIPP Appendix 7, page 213]

We have suggested a solution. Other preservation literature mentions only two possibilities for making complex information durably intelligible: "transformative migration" and "preservation emulation"—and that both approaches are irretrievably flawed. We've begun to see articles that suggest giving up looking for a durable encoding, such as that of Waters in [CLIR 107]. Such thinking assumes that two failed methods demonstrate that no method will work—that nothing new will be found—bba curious premise.

We do not suggest that providing long-term digital preservation problem is simply a technical challenge. For instance, full success would hinge on agreements of the parties involved in generating new technologies and those creating new types of data. However, such agreements cannot be achieved before supportive technical knowledge is available and broadly appreciated. This *Trustworthy 100-Year Digital Object* series tries to articulate the essential technical knowledge, leaving social, legal, and organizational aspects to other works.

Our technical discussion has been as brief as possible, since there exist excellent accounts of general preservation and digital library requirements, of the state of the art, of the viewpoints of archiving institutions, and of practical technical details. [CLIR 107] is an up-to-date assessment of archiving know-how and challenges. [Okerson] presents research library perspectives thoughtfully. [Gilheany] is an excellent tutorial of practical details that include an argument for "acceptable loss"—not seeking perfect information recoverability. For analog information, acceptable loss is practical; for digital information, this compromise can be avoided.

The current article treats only relatively simple digital objects—static data files and the class of programs called filters. Static data files and filters are sufficient for a large fraction of the resources at risk. If it is truly urgent to start preservation activities because significant cultural losses would otherwise occur, our treatment is **a practical basis for implementation of early large-scale repositories.**

§4.3 and §4.4 suggest how more complex bit-string aspects—input and output, timing dependencies and interruptions, and multiprocessing—might be handled. Since we are not yet confident that we know the full set of complexities that should be treated, and have not yet considered efficient ways of handling them practically, we defer their treatment and also suggest that other people might enjoy handling the implicit engineering challenges. This is reasonable because nobody has identified which complex information classes beg for urgent treatment and because the infrastructure required even for simple digital objects is not in place and is likely to require 1-2 years to establish after institutions make a start.

One might argue that the mechanism proposed is more difficult than required, because scholars in centuries to come would have available not only the works we specifically preserve, but also many works that have not been so preserved—an information collection that can be used to guide interpretation that does not depend on our mechanism. For much content, such an argument has merit; it's a statement that digital archeology will often work satisfactorily, even if accomplishing it might be tedious. However, the argument breaks down for cases in which the correctness of some critical fact cannot reliably be determined from available context, as often occurs in computer programs.

Neither the technology sketched above nor the discussion below says anything about semantics—what preserved information means; for that, see [#5]. Furthermore, the current article says nothing about authenticity, except for the obvious fact that there is little point in the authenticity of information that the future consumer cannot exploit. It deals only with the digital communication; however, communication often includes analog steps (see [#9], Figure 1]). Minimizing analog signal errors, such as audio noise and video color distortion,is a topic for audio and video engineers, about whose refined techniques we have no useful comment.

Since **durable encoding** has not yet received much attention in the preservation community, no critical reviews are known to us. We hope that the scientific community will examine the *Trustworthy 100-Year Digital Objects* articles, and identify any functional shortfalls or incorrect reasoning.

## 4.1  Why Durable Encoding Will Work

The core of durable encoding is expressed in the following propositions:

- Any information that can be preserved can be represented by a bit-string, because preserved information must be fixed, as required by copyright law, by archival record principles, and for copying.
- Original undisturbed data are saved as part of preservation packages. This is optional if one chooses to base their UVC interpreter on a transformed representation.
- UVC translators are used only for content portions for which sufficiently simple and reliable EDP standards are not available.
- A well-defined UVC can accomplish any computable transformation from one file to another file.
- An adequate UVC can be specified completely, accurately, and unambiguously. UVC programs need not be very efficient, because they will be executed relatively infrequently, and on machines that will be much faster than today's models.
- The UVC definition and other "bootstrapping" starting points are specified in combinations of natural language, mathematical languages, and a few solid, widely used standards.
- Other preservation infrastructure will include everything needed beyond UVC translators. This infrastructure includes networks of repositories with trust scaffolding, and standards such as Unicode, XML, RDF, and Namespaces. [#3]
- Referential integrity within and among TDOs—references needed to bind to infrastructure—will be protected as described in [#3].
- The difficult task of describing information creation and usage environments is eliminated.
- Nothing specified for durable encoding interferes with other aspects of digital preservation.
- What differs for different data types occurs mostly in 2005 tasks. Available conversion aids are useful "as is".

## 4.2 Efficiency and Skills Needed

How much work will be needed to manage an information corpus? How does the cost of durable encoding compare to what the other methods would cost if they were viable? Whose efforts will be needed, what must they do, and what skills will they need?

Insufficient information is available for precise estimates. The cost of writing a program is difficult to compare to that for other programs. The comparative cost of executions is unknown, partly because of possible optimizations such as batching. Nevertheless, we can estimate crudely by counting how many programs must be written and how many executions might be needed for idealized assumptions.

Suppose that each of **p** data types is represented by **q** instances that occur in **m** M2005 computer types, and that these are to be made available in any of **n** M2105 computer types. Each data instance occurs on at least one M2005 and might be wanted on every M2105. Notice that we count **Case 4** programming languages together with **Case 3** data types.

**Durable encoding:** We need to emulate each M2005 or data interpreter by a UVC program and to emulate the UVC by a M2105 program—**(m+n)** language translators altogether.

**Preservation emulation:** We would need to emulate each M2005 on each M2105—**(m%n)** language translators altogether.

**Transformative migration:** Suppose that between 2005 and 2105, we are forced to migrate the programs **k** times, and that, in each intermediate machine generation, we use a single machine type to preserve the programs. We would have to write a translator for every data type to perform each migration, and to transform every data file each time—**p%(m+n+k-2)** transformation definitions.

**Table 1: Rough cost comparisons**

| Kind of work | Count for transformative migration | Count for preservation emulation | Count for durable encoding |
| --- | --- | --- | --- |
| Writing emulators | 0 | m % n | m + n |
| Writing UVC programs | 0 | 0 | p |
| Writing transformation programs | p x (m+n+k-2) | 0 | 0 |
| Executing periodic migrations | (k-1) x p x q | 0 | 0 |
| Executing in 2105 | p x q | p x q | p x q |

Table 1, together with the estimate that emulators will usually cost more than UVC programs, suggests that durable encoding is more efficient than transformative migration and about as efficient as preservation emulation.

The biggest durable encoding burden falls on (agents for) information producers. A smaller burden falls on (agents for) information consumers. Digital repository custodians acquire no new duties for preservation. The TDO scheme automates tasks that might otherwise be assigned to people. Producers' agents need to **know** today's formats, and to **guess** what end users will understand and what authors intend to communicate. [#5].

For each content data type encountered in **Case 3**—such as Lotus 1-2-3 spreadsheets, PDF files, videos, and statistical tables, some 2005 programmer needs to provide a UVC program for 2105 execution. Each document producer must refer to this program in TDOs he prepares. For each **M2005** whose programs will be preserved, some 2005 programmer must provide an emulator in UVC code.

For each **M2105,** some 2105 programmer needs to provide both a UVC emulator and a restore application that acts as the main program calling and providing input/output services to UVC programs. If a 2105 programmer finds an interpreter of 2005 content too slow, he can write an equivalent compiler. Whoever wants to access preserved content must link these programs and the content of interest.

To make better cost estimates than those above, we need instance count estimates for data types that archiving institutions propose to preserve. Our informal inquiries of archiving institutions have, however, found only very few with careful requirements statements.

## 4.3 Input/Output and Other Advanced Topics

The only programs treated properly in this article are filters—programs that each accept a finite number of finite length input strings and produce a finite number of output strings. We have not worked out treatments for **Case 5** (§3) sufficiently for confident exposition. However, we can speculate about them.

The output of modern program translators (compilers and interpreters) includes unbound names for environmental resources to be resolved during program loading or later during program execution. Their schema are specified for each computing platform supported by the translator. We need to find these specifications and to devise a durably intelligible specification for resources that the Restore Application (Figure 5) needs to provide.

## 4.4 A Pilot Installation and Work Still Needed

The current paper does not propose a particular choice of UVC or of other information that must be taught to our descendants, other than the set(s) of files that are being preserved. To do so would be premature until we have tested with enough examples, built a persuasive prototype, and devised a mechanism for providing execution resource binding (§**Error! Reference source not found.**).

The Koninklijke Bibliotheek (KB, the Dutch National LIbrary) recently deployed a deposit system for electronic publications, based on a system developed by IBM (the Digital Information and Archiving System, DIAS) and the preliminary UVC definition described by [Lorie]. In that framework, IBM and the KB conducted a joint proof of concept study on the UVC. The system, now operational, uses the UVC to decode JPEG files. In addition, [KB] shows how the UVC was used to archive information from a PDF file.

We must still work out careful designs for non-deterministic applications [Ronsse], time-sensitive applications and simulations (**Case 5**). Practical **Case 3** and **Case 4** pilots are still needed for the usual reasons—testing the soundness of ideas, exposing errors and oversights, and carefully estimating costs and the operational skills required. Furthermore, we need to inspect and assess announcements of prominent office automation vendors that their offerings will produce XML files directly, as this potentially makes UVC use unnecessary for very large numbers of documents. Finally, we need to position the current work relative to emerging standards directed at digital preservation [Waibel] and file format information. [Darlington], [JHOVE]]

Ontologies being developed and RDF [RDF] can probably convey meaning as clearly as is theoretically possible, and can be applied to express anything expressible with mathematical graphs. If this opinion is correct, we can devise specific methods that convey as much meaning as can be conveyed. [#5] addresses this topic.

An open question is, "What information renderings do we want to convey to our descendants?" For relatively simple data, such as organizational hierarchies, photographs, legislative text, or simple scientific tables, the answers are often obvious. Other cases are more difficult, partly because they involve questions of authenticity. Deciding optimal action is sufficiently important and subtle to merit careful treatment.

Communicating without contaminating intended content with accidental content seems to be impossible. For instance, I cannot speak without using some voice pitch, and any paper document has paper and ink colors that are usually irrelevant to its meaning. This suggests two questions of broad interest.

➢ Given some information to be preserved, what is the best that can be accomplished towards communicating which encoded information is intended and which is accidental?
➢ Recognizing that few people will understand or be able to generate UVC programs (or care to take the trouble), what tools can we provide "ordinary" users to hide the complexities?

## 5 CONCLUSIONS

Long-term usefulness of information saved today is often imperiled by technological obsolescence that includes replacement of encoding schemes. To overcome this obstacle, we usually need details of today's information practices—details that will probably be more difficult to extract and use years from now than they are today. Prior attempts to bridge over technology changes fail because they save data with dependencies on today's platforms. To isolate, to specify correctly, and to communicate platform details is not only difficult, but also an irrelevant distraction.

For simple digital objects, digital standards that do not depend on ephemeral technology suffice, since they can be specified precisely and intelligibly. For complex digital objects, more is needed. We have described a viable solution. **Durable encoding** uses current hardware and software information in re-write routines whose outputs exclude irrelevant information from preservation bit-strings.

This is accomplished by creating Universal Virtual Computer (UVC) programs that accompany today's content to render it for our descendants. The UVC definition is simple enough that a complete specification can be written to be correctly understood whenever needed. Every step of creating and emulating UVC programs can be executed by programmers of ordinary competence.

Part of each preservation package can be today's bit-strings. No bit need be discarded. No detail should be altered. Durable encoding will not interfere with any invention for exploiting the saved information in new ways (re-purposing), because all essential details about today's context can be saved.

**Durable encoding** must be combined with methods for managing document collections, digital repositories, and integrity and authenticity. **TDO** (**trustworthy digital object**) packaging includes economical solutions for all this. Combining **TDO** packaging with a digital library offering will address all articulated requirements for digital archiving. We believe that **our full proposal provides everything that technology can provide for long-term digital preservation**. We are writing a complete architecture for this digital preservation proposal.

In contrast to the approach sought in most articles addressing digital preservation, our solution makes no new requirements of digital repository technology. **Existing content management (a.k.a. digital library) offerings are adequate, or almost adequate.**

*Trustworthy 100-Year Digital Documents: Evidence Even After Every Witness is Dead* [#3] and the current article together provide enough detail so that **any competent software engineer could implement trustworthy and durable encoding for many digital collections**, safely deferring treating document instances that include the complex data types deferred (§**Error! Reference source not found.**). Most of the technology needed is readily available today either as open-source software or as commercial-off-the-shelf (COTS) offerings. Implementations will consist mostly of integration of such offerings to accommodate needs that vary among users that include repository institutions. There is **no technical impediment to early deployments.**

## Acknowledgements

Robert Morris, sometime Director of the IBM Almaden Research Center, suggested the problem to Raymond Lorie. His encouragements and those of Robin Williams are gratefully acknowledged.

An e-mail dialectic with Reagan Moore suggested explanations used above. Drafts of this article were carefully criticized also by David Bearman and Jennifer Trant, John Bennett, Jim Gray, Peter Lucas, Richard Hess, and John Swinden. Their questions and suggestions are reflected in the current version.